\documentclass{aastex}
\usepackage{emulateapj5}
\begin{document}
\title{Masses, tidal radii and escape speeds in dwarf spheroidal galaxies
under MOND and dark halos compared}
\author{F.~J.~S\'anchez-Salcedo
and X.~Hernandez}
\affil{ Instituto de Astronom\'{\i}a, Universidad Nacional
Aut\'onoma de M\'exico, Ciudad Universitaria,
Apt.~Postal 70 264, C.P. 04510, Mexico City, Mexico}
\email{jsanchez@astroscu.unam.mx,  xavier@astroscu.unam.mx}
\begin{abstract}
We investigate the success and problems of MOdified Newtonian
Dynamics (MOND) in explaining the structural parameters and dynamics
of remote Galactic globular clusters (GCs) and dwarf 
spheroidal (dSph) galaxies.
Using the MOND value for the mass of the Milky Way as inferred from 
the Galactic rotation curve, we derive the tidal radii of Galactic GCs,
and compare to observed values.
Except for Pal 14, the predicted tidal radii of GCs are systematically
larger than the observed nominal values. However, the associated
uncertainties are so large that tidal radii are consistent on the
$1\sigma$ level. We have considered the importance of the
Galactic tidal force on the survival of dSphs under MOND. 
Assuming mass-to-light ratios compatible with a naked stellar population,
we found that the present Galactic dSphs preserve their integrity
over one Hubble time, except Sextans which may survive the
tidal interaction only for several Gyr. 
Mass-to-light ratios as inferred from the internal 
kinematics of dSph galaxies can be used, but they appear  
too large to be accounted for only by
the stellar population in Willman 1, Coma Berenice,
Ursa Minor, Draco, Ursa Major and 
possibly Bo\"{o}tes dwarves. 
Finally, the ability of the Sculptor dwarf to retain the observed population
of low-mass X-ray binaries (LMXBs) is examined. 
Under the MOND paradigm, we find that the retention fraction in 
Sculptor is likely not larger than a few percent.  
Compared to the dark matter scenario, it turns out that 
MOND makes the retention problem worse.
We propose that measurements of the radial velocities
of the observed LMXBs in Sculptor could provide a way to distinguish
between modified gravities or extended and massive dark matter halos.
 
\end{abstract}


\section{Introduction}
The dynamical mass-to-light ratios, $M/L$, derived in galaxies are usually
larger than the expected mass-to-light ratio of the stellar component, 
usually interpreted as
indicating that they must contain dark matter.
Alternatively, one could argue that the discrepancy between total 
mass and baryonic mass is telling us that the Newtonian law of 
gravity is not governing the
dynamics. In particular, the MOdified Newtonian Dynamics (MOND) proposed
by \citet{mil83} has been proven to be successful in explaining how
galaxies rotate, without any dark matter (see Sanders \&
McGaugh 2002 for a review). 
It is important to note that MOND ideas and variants have so far been
introduced, tailored and calibrated
with the explicit objective of accounting for the first order
gravitational effects of hypothetical dark mater
halos: rotation curves in large galaxies and stellar velocity
dispersions in spheroidal systems. One should
also note that a wide range of other gravitational effects
appear in galactic dynamics, which in
principle, offer independent restrictions on any proposed modified
dynamic scheme. Some examples include the gravitational stability of galactic 
disks (e.g., S\'anchez-Salcedo \& Hidalgo-G\'amez 1999) or relaxation
processes in stellar systems \citep{cio04}. Concerning
dSphs, in S\'anchez-Salcedo et al.~(2006) we investigated the effects
the enhanced gravitational relevance of
the stellar population under MOND would have upon the problem of
the orbital decay of globular clusters (GCs) due to
dynamical friction,
and found that although such a proposal can account for the
velocity dispersion measurements, in-spiraling
times become uncomfortably short. 

\citet{ger92} and \citet{ger94}
derived the mass-to-light ratios in MOND
for seven dwarfs from virial arguments.
Later on, \citet{mil95} reanalyzed the
$M/L$ values with new data. He concluded that when uncertainties in
the observed luminosity and central velocity dispersion are included,
the values agree with those expected for a normal old stellar population.
The flux of data about the kinematics of Local dSphs is growing rapidly.  
The new determinations of line-of-sight 
velocity profiles, the recent discovery
of a dozen dSph galaxies, and the detection of low-mass
X-ray binaries (LMXBs) in Sculptor may give crucial clues on the
dark matter problem and its alternatives.
In the Newtonian approach, these observations lend support to the 
cosmological-motivated interpretation that some dSphs reside
within the most massive and extended substructure dark matter halos 
\citep{kle01,kle05,deh06}.
The fact that some dSph galaxies possess two distinct populations, 
as inferred from studies of resolved stellar populations, e.g.~\citet{her00}
for Carina and Leo I,
also suggests an extended dark matter halo in
order to prevent the loss of all the gas during the first starburst.
The question that arises is whether MOND can mimic an extended
halo in dSphs or not.
\citet{lok02} and \citet{lok06} have noticed that MOND 
experiences a serious difficulty in explaining the velocity dispersion 
profile of Draco without dark matter. 

A wide view of the successes and problems
of MOND in dSph galaxies in comparison to the dark matter scenario 
can motivate new tests on the nature of dark matter and its
alternatives.  Here we will investigate the ability of MOND in
explaining the observed tidal radii in Galactic GCs
and dSph galaxies (\S 2), and the problem of the retention of LMXBs 
in dSphs (\S 3), within the alternative optics of MOND without dark matter.
As a byproduct, an updated compilation of the mass-to-light ratios in 
satellite dSphs under MOND is given.

\section{Tidal radii of GCs and the survival of dSphs}
\label{sec:tides}
In any law of gravity, the global $M/L$ ratio in a spherical system
can be derived in different ways;
either modeling the observed velocity dispersion
profile of the stars or using the observed King tidal radius.
The observed tidal radii can provide lower limits 
for the $M/L$ that are independent of kinematic data
(e.g., Faber \& Lin 1983; Oh, Lin \& Aarseth 1995; Pryor 1996; 
Burkert 1997; Walcher et al.~2003).

The dynamical \citet{kin62} tidal radius of a GC or a dSph orbiting in a
circular orbit of radius $D$ around a galaxy with a logarithmic 
potential is defined as:
\begin{equation}
r_{t}=k\left[\frac{M_{dw}}{2M_{G}(D)}\right]^{1/3}D,
\label{eq:kingtidal}
\end{equation}
where $M_{dw}$ is the mass of the satellite, $M_{G}(D)$
the host galaxy mass inside $D$ and $k$ is Keenan's (1981a,b) factor,
which depends on the gravity law,
and accounts for the elongation of the zero-velocity surface along the line
joining the cluster and galactic centers.
For eccentric orbits,
\begin{equation}
r_{t}=ka \left[\frac{M_{dw}}{M_{G}(a)}\right]^{1/3}
\left[\frac{(1-e)^{2}}{\left[(1+e)^{2}/2e\right]\ln\left[(1+e)/(1-e)\right]+1}
\right]^{1/3},
\label{eq:kingtidalecc}
\end{equation}
where $e$ and $a$ are the orbital eccentricity and the semi-major axis,
respectively.  This formula for the tidal radius 
is valid for both Newtonian (e.g., Oh, Lin \& Aarseth 1992) 
and MOND gravities \citep{ger92,bau05,zha06}
provided that the masses and the
Keenan factor are taken appropriately.
In the MONDian case without dark matter, $M_{dw}$ and $M_{G}$ are
the ``true'' baryonic masses of the dSph satellite and the host 
galaxy, respectively.
Knowing the Galactic mass radial profile and the eccentricity $e$ and by
equating King's tidal radius, $r_{t}$, with 
the observed limiting radius, a lower limit on $M/L$ of the satellite
system can be inferred.

The nominal value $k=1$, as first suggested in King's paper (1962),
corresponding to the distance of the last closed zero-velocity surface
along the line joining the center of the galaxy with the center of the
satellite system, 
is commonly adopted as a useful fiducial length scale under Newtonian
gravity, after the work of
\citet{oh92} and \citet{oh95}, who
demonstrated that even in the absence of significant two-body
relaxation processes, the density profile of GCs in circular
orbit resembles the King-Michie
model with an asymptotic limiting radius consistent with the tidal
radius as defined by \citet{kin62}, with $k=1$. 
If we adopt the same definition of tidal radius in MOND, i.e.~the
distance to the zero-velocity surface along the axis pointing
in the radial direction, we also get $k=1$ \citep{zha06}.

The parameter $F$, defined as the ratio between the observed cutoff
radius and the dynamical tidal radius as defined in
Eq.~(\ref{eq:kingtidalecc})
with $k=1$, is a useful indicator of the importance of the
Galactic tidal force on the internal structure of GCs
and dSphs \citep{oh95, pia95}. 
For values $F<1$, the tidal force is unable to stretch
the satellite object significantly, and only internal processes
may be considered.  In this case,
the stellar component of the dSph galaxies is not tidally truncated,
and mass estimators based on the tidal radius underestimate 
the total $M/L$ substantially. 
At values $F\sim 2$, a dSph can survive the
tidal interaction for several Gyr. 
At these values a dSph could actually be
unbound but not yet dispersed, as the timescales
for disruption exceed several Gyrs. With the effect of the force
increasing rapidly with increasing $F$, for $F\sim 3$ the satellite
disintegrates in a few orbits. Since it is unlikely that we are observing
satellites in such a short-lived evolutionary state (except Sagittarius),
$F<2$ is expected.  

The values of $F$ under MOND, $F_{M}$, for Galactic GCs and satellite dSphs 
have been calculated recently by \citet{zha05}.
\citet{zha05} uses a Keenan factor $k=\sqrt{2}/3$ 
and equates the mass ratio with the luminosity ratio, 
i.e.~$M_{dw}/M_{G}\approx L_{dw}/L_{G}$ in Eq.~(\ref{eq:kingtidal}),
where $L_{dw}$ and $L_{G}$ are the luminosities of the dwarf galaxy
and our Galaxy, respectively. 
The large scatter found in the values of $F_{M}$ was interpreted 
by \citet{zha05} as challenging for any baryonic MOND Universe.

In this Section, we revisit the consistency of the predictions
of MOND with observations using the tidal radius theory, 
assuming the mass of the Milky Way to be known. 
We discuss GCs (\S \ref{sec:globularclusters})
and dSphs (\S \ref{sec:dsphs}) separately, since remote GCs
are likely tidally truncated, whereas dSph galaxies are not 
necessarily tidally truncated, even though their luminosity profiles
can be fitted by King models. 

\subsection{Globular clusters}
\label{sec:globularclusters}

In order to estimate the tidal radii of Galactic globular clusters
under MOND, we must start by adopting a mass model for the Milky Way
under MOND. We take this from \citet{fam05} 
(their model I) who use the
Galactic rotation curve, over the region where it is well measured, the first
10 kpc, to derive a total baryonic mass of 
$M_{G}=0.6$--$0.8 \times 10^{11}$ M$_{\odot}$ (see also Dehnen \& Binney 1998).
This yields a circular rotation at the solar circle of $200$  km s$^{-1}$ 
and good agreement with dynamical measurements at all radii internal to 10 kpc.
Since the asymptotic circular velocity of test particles around
a mass $M_{G}$ is $V=(GM_{G}a_{0})^{1/4}$, this implies an asymptotic circular 
velocity of $170\pm 5$ km s$^{-1}$
(essentially shared by most of their models, see also Famaey et al.~2007), 
for our adopted 
$a_{0}=0.9\times 10^{-8}$ cm s$^{-2}$, which is the value derived
in Begeman et al.~(1991) rescaled to the new distance scale \citep{bot02}.  
This mass compares quite well with the expected baryonic disk mass for a galaxy with 
a disk V band luminosity of $\sim 1.5\times 10^{10}$ L$_{\odot, V}$. 
The asymptotic
value of the rotation curve might seem a little on the low side, but one must
bear in mind that more familiar values of order of $200$  km s$^{-1}$ 
are the result of extrapolating dark halo mass models beyond the 
regions where the rotation curve as such is actually measured.

It is worthwhile to compare the 
King tidal radii in the Newtonian dark matter scenario
and in MOND.  According to Eq.~(\ref{eq:kingtidal}), 
the tidal radius in the conventional dark matter model depends on the
total mass of the host galaxy as
$[M_{G,bar}(D)+M_{G,dm}(D)]^{1/3}$, where $M_{G,bar}$ and $M_{G,dm}$ are
the baryonic and dark mass, respectively.
In MOND, the tidal radius goes as $[M_{bar}(D)]^{1/3}$, 
where $M_{bar}$ is the baryonic mass only. So that, remote GCs (those
at $D>35$ kpc)
should be the best discriminators between MOND and Newtonian dynamics
as the dark halo mass becomes significant at large distances. 

Unfortunately, the range $D>35$ kpc is mainly 
populated by sparse, low-luminosity clusters with rather uncertain
tidal radii and integrated magnitudes.
For illustration, for NGC 2419, which is the fourth most luminous cluster
of the Galaxy, discrepant values for the global $M/L$ and
cutoff radius can be found in the literature. \citet{pry93} report
$M/L=1.2$ and $r_{t}=230$ pc (see also Trager et al.~1995), whereas
Olszewski et al.~(1993) derive a global $M/L$ ratio of $0.7\pm 0.4$ and
$r_{t}=280$ pc. As discussed in detail by 
\citet{wak81}, \citet{inn83} and \citet{bel04}, who
estimate $M_{G}$ under Newtonian dynamics using the tidal radii
of remote GCs, the observed tidal radii are affected by large uncertainties.
The conventional method for determining the tidal radii involves a large
outward extrapolation of the surface brightness profile. Innanen
et al.~(1983) suggest random errors of a factor $\sim 2$ in $r_{t}$. 
Moreover, the 
adoption of different models to fit them may lead to significantly larger
limiting radii than estimated with King's models \citep{mcl03}.
Our first objective is to test MOND using the observed tidal radii of 
remote GCs, including fully all the sources of uncertainty in the analysis.

The complete list of known GCs beyond $35$ kpc until 2006 is
Pal 2, Pal 15, NGC 7006, Pixis, Pal 14, NGC 2419, Eridanus, Pal 3, Pal 4 and
AM-1 \citep{har96}. 
However, Pal 2, Pal 15 and Pyxis are not suited for the present
study because they are strongly affected by extinction. 
Two new extremely low luminosity GCs have been discovered recently
(Koposov et al.~2007), but
the number of stars detected is not enough
to measure precisely their luminosities and tidal radii. 

Those GCs with orbital periods, $P$, much shorter than the half-mass
internal relaxation time, $t_{rh}$, are expected to be limited 
by Galactic tides at perigalacticon, because internal relaxation 
is unable to restore
a larger limiting radius before the next perigalactic passage
(e.g., Oh et al.~1995). In clusters with an orbital period similar or
larger than the relaxation time, internal relaxation is able to
repopulate their external regions after perigalacticon passage and
their tidal radii may be comparable to the instantaneous
theoretical value and, hence, values $F\sim 1$ are expected \citep{bel04}. 
For the remote GCs of our sample from which orbital parameters are available,
NGC 7006 and Pal 3, have $P/t_{rh}\sim 1$ \citep{har96}. Therefore,
we can be certain that they have $F\sim 1$. 
Eridanus and AM-1, situated at a galactocentric distance 
of $\sim 95$ kpc and $123$ kpc, respectively, 
are likely limited by the Galactic tidal force at its present position
because their $t_{rh}$ value $\approx 2.5$--$5$ Gyr is short compared
to their orbital periods.
In the case of NGC 2419, $t_{rh}\approx 35$ Gyr. Hence, it should be 
limited by Galactic tides at perigalacticon. Finally, 
$P/t_{rh}$ is difficult to infer for Pal 14 and Pal 4.
We consider as a working assumption that the observed
tidal radii of distant GCs, except for NGC 2419, are probes of the Galactic
tidal force at their present position, bearing in mind that we may
overestimate $r_{t}$ for Pal 14 and Pal 4. The predicted
$r_{t}$ values are estimated according to Eq.~(\ref{eq:kingtidal}) for all GCs, 
except NGC 2419, in a MOND model where the Galactic mass is fixed by the 
inner rotation curve. 

For NGC 2419 we are forced to assume an orbital eccentricity.
Since its orbital parameters are unknown we will follow a statistical
procedure. For the $48$ GCs with determinations of the orbital parameters,
$e=0.5\pm 0.25$ at the $1\sigma$ level \citep{all06}. Therefore,
we will estimate $r_{t}$ for NGC 2419 using Eq.~(\ref{eq:kingtidalecc})
and an eccentricity taken as a most likely value of 0.5, 
with a $1\sigma$ uncertainty of 0.25. We should note
that there is a trend on the eccentricity with their apogalacticon
distance: most GCs with apogalacticon distances $\lesssim 10$ kpc 
have eccentricities between $0$ and $0.6$, whereas those with
apogalacticons $\gtrsim 10$ kpc the eccentricities are populated mainly
in the interval $0.4$--$1.0$. Therefore, adopting $e=0.5\pm 0.25$
we are overestimating the upper error range on $r_{t}$ since the 
eccentricity of NGC 2419 is likely $>0.4$.

When estimating $r_{t}$ from Eqs.~(\ref{eq:kingtidal}) or
(\ref{eq:kingtidalecc}), $D$
is derived from the distance modulus $\mu_{V}$,
the color excess $E(B-V)$ and the galactic coordinates, whereas
the mass of the cluster is derived from the apparent integrated $V$-magnitude
$V_{t}$, $\mu_{V}$ and the stellar mass-to-light ratio $M/L$.
Reddening correction, though small (less than $0.1$ mag in the
color excess for all of the GCs considered), was taken into account.
As said before, in order to quantify the success or failure of MOND
in explaining the measured tidal radii of GCs, we should include the
associated uncertainties in the analysis:
there are dispersions around $\mu_{V}$,
$V_{t}$, $M/L$ and Keenan's factor.   We will
proceed adopting the same associated uncertainties on input parameters
as \citet{bel04} and are briefly described below.

(i) The mean values of $\mu_{V}$ and $V_{t}$ were extracted from \citet{har96}.
The standard deviations for both $\mu_{V}$ and $V_{t}$ were taken
of $0.1$ mag for
NGC 2419 and NGC 7006 and $0.2$ mag for the remaining clusters.
The errors in $\mu_{V}$ are thought to account for both the
measurement errors and the uncertainties affecting the distance
scale of GCs \citep{cac99,bel04}.
No scatter was associated to the color excess;
it was kept fixed for each GC.

(ii) The $M/L$ values follow a truncated Gaussian distribution with a
mean value of $1.2$ and a standard deviation of $0.4$,
so that $M/L=1.2\pm 0.4$.  The distribution is truncated at $M/L= 0.5$
(e.g., Pryor \& Meylan 1993; Feltzing et al.~1999; Parmentier \& Gilmore 2001).

(iii) The MONDian Keenan factor $k$ is extracted
from a uniform distribution bracketed by the intermediate and
radial Roche radii, i.e.~$0.35<k<1.0$.  In fact, 
detailed analysis of the tidal radii of Galactic GCs in the Newtonian
case\footnote{Note that GCs are supposed to contain no dark matter 
in these studies.} suggest that the Newtonian
$k$ values must vary from cluster-to-cluster in the range $0.5\leq k\leq 1.0$
(e.g., Bellazzini 2004), probably because there are several
mechanisms that induce an outward migration of stars
and the escape of all extra-tidal stars is by no means assured.
Therefore, it turns out that the possible $k$ values are bracketed by
the intermediate and radial Roche radii, which translates into
$0.35<k<1.0$ in MOND.

\begin{figure}
\plotone{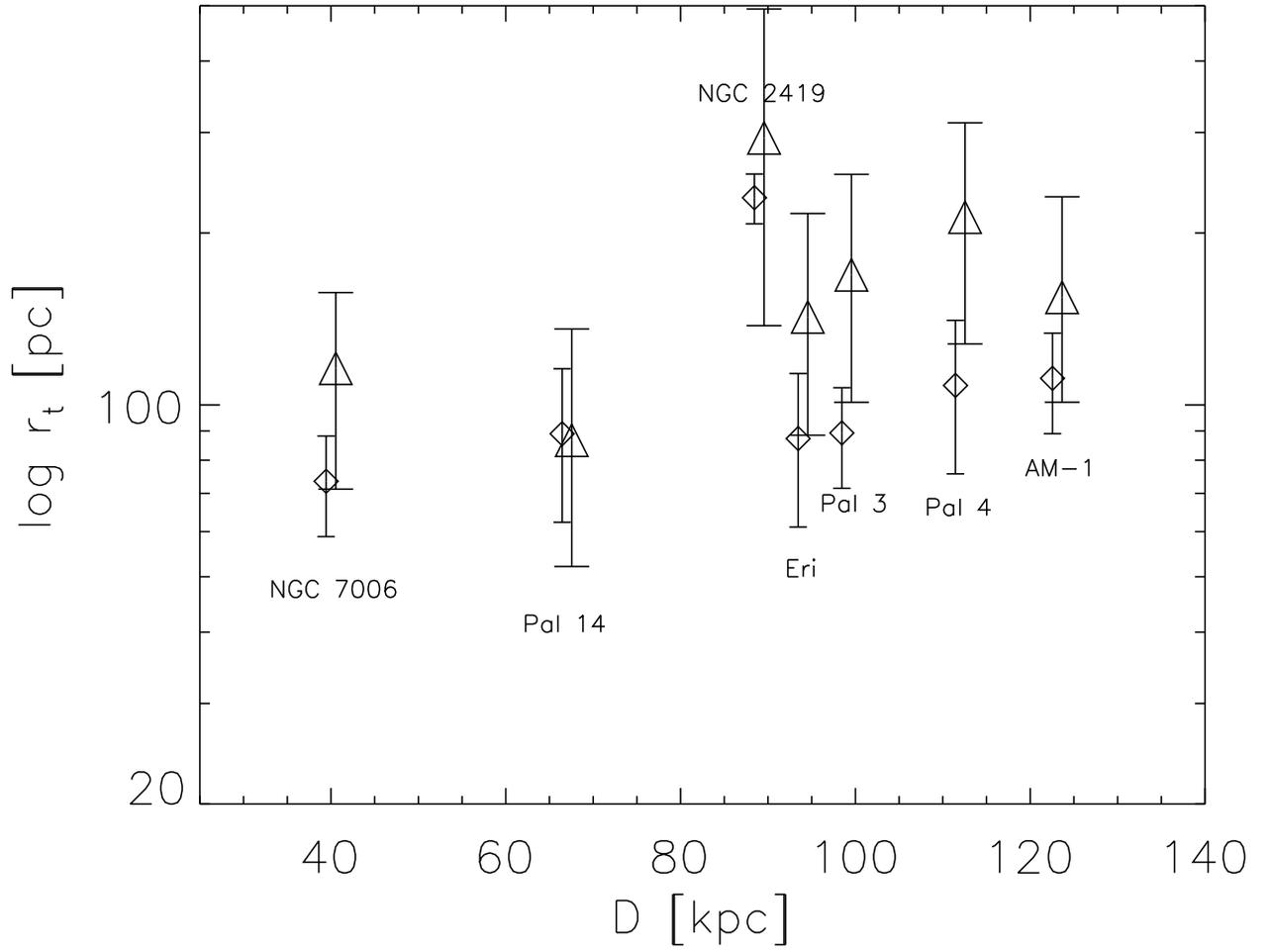}
  \caption{Comparison of tidal radii in MOND with $M_{G}=0.7\times 10^{11}$
M$_{\odot}$ for remote GCs. 
The error bars enclose $1\sigma$ confidence limit range. 
The predicted tidal radii (triangles) are compared with the observed
tidal radii (diamonds). The symbols have been shifted horizontally to make the
plot readable. The observed limiting radii were taken from Harris \& 
van den Bergh (1984) for Pal 14; Harris (1996) for NGC 7006, NGC 2419,
Eri, and Pal 4; from Hilker (2006) for Pal 3 and AM-1. }
\label{fig:bellaMOND}
\end{figure}

The inferred tidal radii with $1\sigma$ error bars 
for $M_{G}=0.7\times 10^{11}$ M$_{\odot}$ 
(see \S \ref{sec:globularclusters})
are shown in Fig.~\ref{fig:bellaMOND}. For comparison, the observed
limiting radii 
are also included. From inspection of the quality and extent of the
surface brightness profile, \citet{bel04} assigns a standard
deviation for the observed $r_{t}$ values of $10\%$ for NGC 2419, 
$20\%$ for NGC 7006 and $30\%$ for the remaining clusters. 
However, estimates of the tidal radii of Pal 3 and AM-1 have been
superseded by \citet{hil06}. Hence we have adopted these updated values
with an uncertainty of $20\%$. It can be seen that 
the central predicted values are systematically larger than the observed values
except for Pal 14. As anticipated, however, the uncertainties are so
large that it is not possible to rule out MOND at all. The most problematic
cases are Pal 3 and Pal 4, for which there is little overlap between 
the error bars.
However, since the observed cutoff radii are plagued with
systematic errors  
(e.g., Heggie \& Ramamani 1995; McLaughlin \& Meylan 2003; Bellazzini 2004), 
not included in the present analysis, 
we conclude that 
the present observational data does not allow us
to distinguish between MOND and dark matter halos. 
If future more refined measurements simply reduce error bars, leaving central inferred values unchanged, 
tidal radii of GCs will indeed rule out MOND.

The differences
between our conclusion and those of \citet{zha05} are probably due to the fact
that we included all relevant sources of uncertainties in the problem, and 
to having modeled each GC using the observed $M/L$ (with their uncertainties),
rather than taking a generic value.

In the following section we shall extend the previous analysis of 
globular clusters to the case of dSph galaxies and evaluate the
importance of the Galactic tide on their structure,
which provide a new test for MOND.

\subsection{Dwarf spheroidal galaxies}
\label{sec:dsphs}
dSph galaxies are not necessarily tidally truncated systems because
the timescale for two-body relaxation, which
is ultimately responsible for the diffusion of stars to the outskirts
of the dSph galaxy, may be so large that internal relaxation
would be unable to smooth any hypothetical
sharp edge in the stellar luminosity profile in one Hubble time.
In the case of the remote Leo II dSph, for instance, the two-body relaxation
timescale is $>300$ Gyr even under modified dynamics 
(derived using Eq.~42 in Ciotti \& Binney 2004 with a ratio between dark matter
and luminous matter of ${\mathcal{R}}\approx 20$)
and, hence, the fractional mass-loss rate 
$M_{dw}^{-1}dM_{dw}/dt\leq -0.045/t_{rh}\approx -1.5\times 10^{-4}$ Gyr$^{-1}$
\citep{gne99}, 
is so small that the photometric contribution of unbound stars 
would be gone in the process of subtraction of the background population.

We have selected all the Milky Way satellites placed at $40<D<260$ kpc
with determinations of their internal dynamics:
strong candidates of the satellite
companions to the Milky Way, namely 
Ursa Major II, Segue 1,
Leo T and Bo\"{o}tes II dwarves are excluded.
Canes Venatici I is also excluded because its complex dynamical
structure would require a more sophisticated analysis (Ibata et al.~2006)
and Leo IV because the kinematic data available is still very poor. 
The observed limiting radii were taken from Irwin \& Hatzidimitriou (1995, 
henceforth IH95).

The predicted tidal radii under MOND theory were derived 
from Eq.~(\ref{eq:kingtidal}) using $M_{G}=0.7\times 10^{11}$ M$_{\odot}$
once $M_{dw}$ is known.
We have estimated $M_{dw}$ in two different manners.
In a MONDian world, the dynamical mass is fully determined by the stars. 
If a value $M/L=3$, corresponding to a normal old stellar population,
is assumed for all the dSphs, then
$F_{M}<1.25$ for all of them except for Sextans,
whose value is $F_{M}=2.1$. This indicates that in order 
to have $F_M<2$ for all galaxies, we should invoke a $M/L\sim 4$. 
Are these values of $M/L$
consistent with their internal kinematics? In order to answer this question,
the MOND $M/L$ values, $(M/L)_{M}$, were inferred from the internal 
dynamics of the stars based on the most recent data.

\citet{mil95} defined a parameter $\eta\equiv 1.5
(\sigma/V)^2(D/r_{c})$,
with $\sigma$ the stellar velocity dispersion of the dwarf, $r_{c}$
its core radius, and $V$ the galactic rotational velocity at $D$, which
coincides with the asymptotic rotation velocity $V_{\infty}$ for
all the dwarfs. When $\eta \gg 1$, the dwarf can be treated as
an isolated system and
the global $M/L$ was obtained exploiting a simple relation between the
total mass $M$ and the mean line-of-sight velocity dispersion, $\sigma$,
in the isotropic, deep MOND case:
\begin{equation}
M=\frac{81}{4}\frac{\sigma^4}{Ga_{0}}
\label{eq:isorel}
\end{equation}
\citep{ger92,mil94}.
When $\eta \ll 1$ and all accelerations relevant to the
dwarf dynamics are smaller than $a_{0}$, the dynamics becomes quasi-Newtonian.
The MOND $M/L$ for an object in the quasi-Newtonian
regime can be obtained immediately from its Newtonian estimate,
\begin{equation}
\left(\frac{M}{L}\right)_{M}=\left(\frac{g_{\rm ext}}{a_{0}}\right)
\left(\frac{M}{L}\right)_{N},
\label{eq:qnrel}
\end{equation}
where $g_{\rm ext}$ is the acceleration of the system in the external field.
We must notice, however, that due to the fact that the acceleration 
becomes anisotropic (Appendix \ref{sec:A1}), the above relation  
is not strictly exact. Some idea of the uncertainty can be gained from the
case of a homogeneous sphere, which we derive in Appendix \ref{sec:B1}. 
The mass-to-light ratio in that
ideal case is underestimated by a fraction $\sim 30\%$ 
(Appendix \ref{sec:B1}) and, hence, due to the uncertainties in the structural 
parameters of these galaxies, this error is unimportant.
For an isolated spherical system in the deep MOND limit, 
the central, MOND $M/L$ value,
$(M/L)_{0,M}$, as a function of the Newtonian estimator $(M/L)_{0,N}$
is given by: 
\begin{equation}
\left(\frac{M}{L}\right)_{0,M}=1.23\left(\frac{\sigma_{0}^{2}}{a_{0}r_{c}}
\right) \left(\frac{M}{L}\right)_{0,N}, 
\label{eq:centralMOND}
\end{equation}
with $\sigma_{0}$ the central velocity dispersion and $r_{c}$ the core radius
\citep{ger92}.

In Appendix \ref{sec:D1} we briefly describe the observational parameters
used to infer the mass-to-light ratios for each galaxy individually.
Bo\"{o}tes, Sculptor, Sextans and Hercules have been assumed to be in the 
quasi-linear Newtonian regime, whereas the rest of the dSph were treated 
in the isolated limit.   
The derived global $(M/L)_{N}$ and $(M/L)_{M}$ are listed, when available, 
in Table \ref{table:parameters}, with $a_{0}=0.9\times 10^{-8}$ cm s$^{-2}$
(Bottema et al.~2002; see \S \ref{sec:globularclusters}) and
the updated luminosities.
For Coma Berenices, Bo\"{o}tes, Ursa Major, Hercules
and Canes Venaatici II, the values reported correspond 
to the central mass-to-light ratios.
When the derived $(M/L)_{M}$ ratios lie within the range
$0.5<(M/L)_{M}<12$, they might be
in agreement with a `naked' old stellar population
\citep{que95,rom03}.
Only when the resulting value is well out this range, we provide a lower
limit at the $2\sigma$ confidence level,
assuming that dSph galaxies have a isotropic velocity
distribution; corrections for this are of as much as a factor $2$.
We wish to know if when assuming these lower $2\sigma$ limits on $(M/L)_{M}$,
it also holds that $F_{M}<2$. However,
the reader interested in the mean $(M/L)_{M}$ values is referred to
Appendix \ref{sec:D1}.

The observed values of the limiting radii inferred by IH95
were used to derive $F_{N}$ and $F_{M}$ since they 
are usually adopted as the nominal values. 
The observed $r_{t}$ in a certain galaxy
may depend on the type of stars selected for its determination.
For instance, red giant branch stars (RGB) are
significantly less centrally condensated than the blue horizontal
branch population in Sculptor (Westfall et al.~2006). More importantly,
recent studies in some dSphs find a ``break'' in the
King profile. The population beyond the break radius must signify
either the presence of an extremely broad distribution of bound stars
or the presence of unbound tidal debris (e.g., Westfall et al.~2006).
In the latter case, $r_{t}$ should be identified with the
break radius.  For Ursa Minor, Sculptor, Draco, Fornax and Leo I,
we have listed additional rows in Table \ref{table:parameters}
with new determinations of $r_{t}$ and luminosity.  For conciseness,
however, we do not give all the possible combinations of $r_{t}$
and $L$ but only
the most relevant to assess the influence of observational uncertainties.

The derived values of $F_{M}$ suggest that
Leo I is the least likely to be affected by tidal effects.
On the opposite side, Ursa Minor, Sculptor and Sextans 
may have $F_{M}\gtrsim 1$. 
Is there any evidence of the tidal distortion in these galaxies?

From photometric data of Ursa Minor, Palma et al.~(2003) suggest that
this object is evolving significantly because of the tidal influence
of the Milky Way, although they cannot determine if the extra-tidal
stars are now really unbound.
In the case of Sculptor, the resultant value $F_{M}=0.8$--$1.1$
is in perfect agreement with its appearance 
(see Section \ref{sec:sculptormain}).
We turn now our attention to Sextans. 
Applying Eq.~(\ref{eq:kingtidal}) with $k=1$, $M/L=12$, $D=86$ kpc
and $M_{G}=0.7\times 10^{11}$
M$_{\odot}$, we obtain $r_{t}=2.15$ kpc. This value should
be compared with that reported by IH95
(Table \ref{table:parameters}) to obtain $F_{M}=1.3\pm 0.5$,
compatible with the requirement $F\leq 2$.
We suggest that deep observations of Sextans, both photometrically
and spectroscopically, out to radii larger than those
where tails are detectable, may provide a test of MOND. 
If it is confirmed that Sextans is not being tidally distorted,
this galaxy may need an extended halo of dark matter even under MOND dynamics.
If so, the appeal of MOND would appear to be significantly weakened.
In the standard Newtonian dark matter scenario $F_{N}=1.4$ in Sextans,
but it is likely that the global $(M/L)_{N}$ is larger than the quoted
value of $\sim 100$ (Kleyna et al.~2004).

Finally, Draco, which has a large kinematic
$(M/L)_{M}$, would survive the Galactic tide even if a $(M/L)_{M}=2$, 
typical of a old stellar population, was adopted.

We conclude that it is very difficult at present to use the tidal radii
of dSphs to distinguish between MOND and standard CDM. 
Until kinematic measurements definitively identify the tidal radii,
the tidal approach should be treated with caution, as it is only
an indicator of the importance of the Galactic tidal force on the
structure. 
In MOND, the dynamical $M/L$ are systematically larger than the $M/L$
inferred using the tidal radii (i.e.~the requirement $F_{M}<2$). Therefore, 
explaining the internal dynamics is a more profound question
than the tidal radii. In fact,
the high $(M/L)_{M}\gtrsim 20$ values required in Willman 1, Coma Berenice,
Ursa Minor, Draco, Ursa Major and possibly Bo\"{o}tes
are difficult to accept if MOND is a valid
alternative to dark matter in dSph galaxies. 
Willman 1, Coma Berenice and Ursa Major are unique targets to test the success
or failure of modified gravities at galactic scales.
One could argue that an appropriate
choice of anisotropy in the orbits of stars within these galaxies
could reduce somehow these estimates. However, \citet{lok02}   
found the opposite trend in Draco and Fornax.

\begin{figure}
\plotone{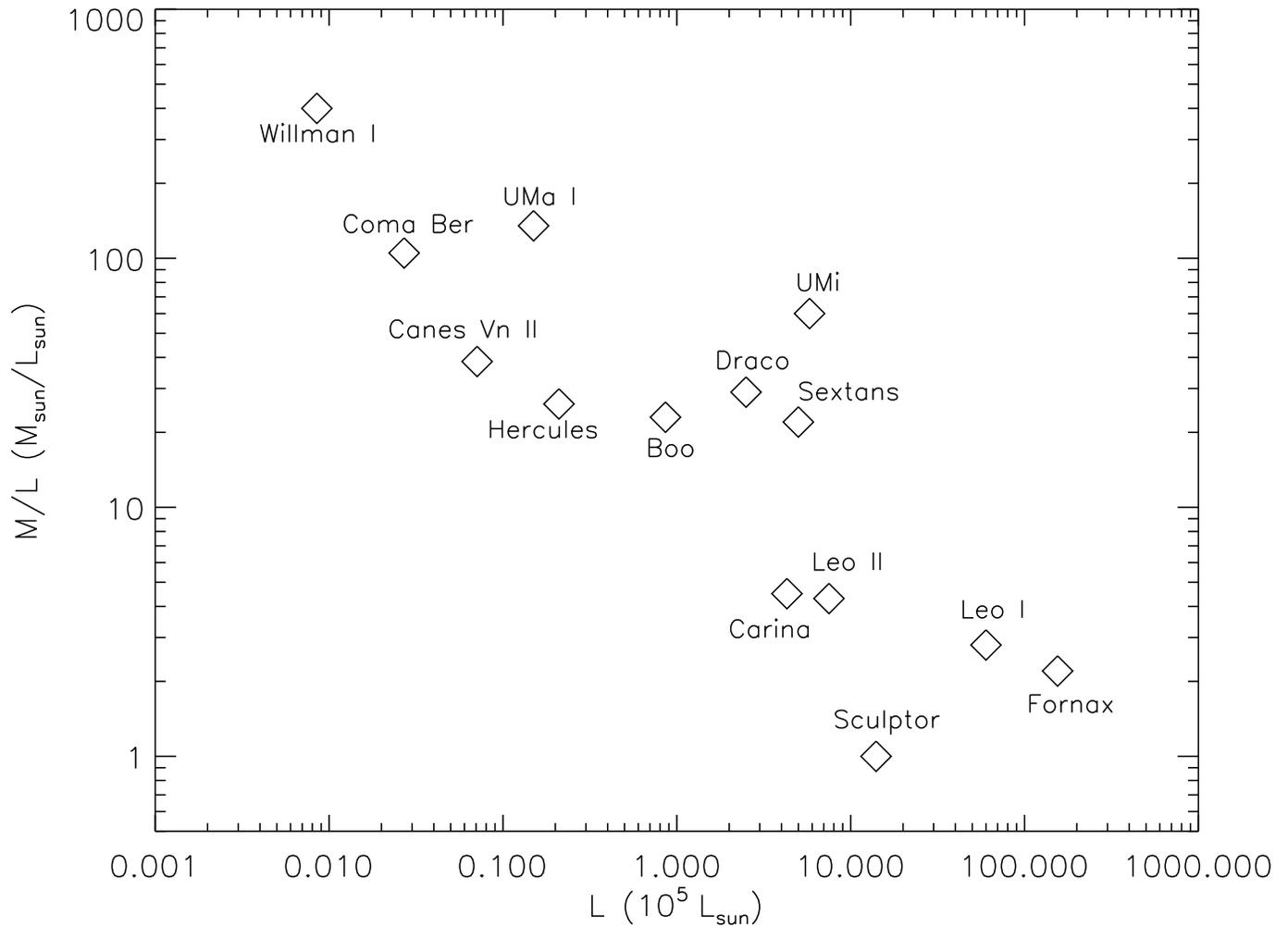}
  \caption{Inferred MONDian mass-to-light ratios versus luminosity for
the known Galactic dSph galaxies at $40<D<260$ kpc and with
velocity dispersion estimates (see text).}
  \label{fig:mll}
\end{figure}

The relation between $M/L$ and luminosity 
is shown in Figure \ref{fig:mll} for our sample of galaxies.
This figure is based on that given in Mateo et al.~(1998) but for MOND.
Even with considerable scatter, the clear trend between $M/L$ and
luminosity is because MOND is based on a characteristic
acceleration and dSph galaxies present a characteristic velocity
dispersion. According to this relation, the kinematics of the faintest dSph
galaxies will be puzzling under MOND.

Galactic tides could, in principle, affect the kinematics and the
apparent $M/L$ derived from kinematic studies that assume dynamical
equilibrium. For instance, \citet{kuh89} and \citet{kro97}
presented tidal models
to explain large values of $M/L$ in dwarf galaxies
without restoring to dark matter. However, there is growing evidence
that galactic tides cannot inflate the global $M/L$ values: tides
produce large ordered motions rather than large random motions \citep{pia95}.
Oh et al.~(1995) found that the unbound but not yet dispersed
systems have velocity dispersions that are comparable to the virial
equilibrium value prior to disruption.
Therefore, if the measurements do not seriously overestimate the velocity
dispersion, either the mentioned galaxies are not in dynamical equilibrium 
due to recent gas mass loss or to a major merger with another dwarf
galaxy, undergoing still a relaxation process, or they 
contain large amounts of dark matter even under modified dynamics. 
We believe that it is unlikely that the gas has been stripped from 
these galaxies 
only very recently, in the last few internal crossing-times.
In particular, the fact that Draco and Ursa Minor have only very old 
stellar populations
($\gtrsim 10$ Gyr), suggests that they have possessed a 
dynamically-negligible fraction of mass in gas since a long time ago
(Mayer et al.~2007).

\section{Retention of low-mass X-ray binaries}
Recently, \citet{deh06} have suggested that 
in order for Sculptor to retain the population of low-mass X-ray binaries
observed by Maccarone et al.~(2005) 
an extended dark matter halo of $\geq 10^9$ M$_{\odot}$ is required,
i.e.~$(M/L)_{N}\sim 600$.
Here we explore the implications of this observation in order to  
see if the stronger gravity due to the stars alone under MOND also has the 
ability of retaining the observed LMXBs or not.

\subsection{The Sculptor dwarf spheroidal: Internal parameters 
and proper motion}
\label{sec:sculptormain}
The Sculptor dSph is a satellite galaxy of the Milky Way, which
presents a luminosity $(1.4\pm 0.6)\times 10^{6}$ L$_{\odot}$
(IH95). According to the proper motion
measurements by Piatek et al.~(2006), it is on a polar orbit with
apogalacticon at $122$ kpc and perigalacticon at $68$ kpc.
The current galactocentric distance of Sculptor is $79\pm 4$ kpc
\citep{mat98}. A King model profile of limiting radius $79'.6\pm 3'.3$,
corresponding to $1.85$ kpc and core radius $7'.14\pm 0.'33$ ($165$ pc),
fits the density profile well out to $60'$. Beyond this radius,
a break population, which extends up to $150'$, has been found
(Westfall et al.~2006).
The ellipticity of the isodensity contours increases with increasing
projected radius from the center of the dSph, from rounder to a value
$0.32$ in the outermost region (IH95; Westfall et al.~2006).
Since there is no evidence for rotation, this flattening should be due
to anisotropy in the velocity distribution, but it may also indicate
that Sculptor is tidally distorted.

There is evidence that Sculptor contains two distinct ancient (both
$\gtrsim 10$ Gyr old) stellar components. Tolstoy et al.~(2004) derive
different velocity dispersions for Sculptor stars separated into
metallicity groups. The line-of-sight velocity dispersion of the metal-rich component
is $\approx 7\pm 1$ km s$^{-1}$, which is similar to the
central line-of-sight velocity dispersion in
Sculptor $6.2 \pm 1.1$ km s$^{-1}$ (Queloz et al.~1995; Westfall et al.~2006).
Interestingly, the metal-poor component has a velocity dispersion
of $\approx 11 \pm 1$ km s$^{-1}$ (see also Clementini et al.~2005).
Westfall et al.~(2006) derive a global velocity dispersion of $8.8\pm 0.6$
km s$^{-1}$, which is intermediary to those that Tolstoy et al.~(2004) report
for these metallicity-separated populations. 
In the following, we will take the
latter value as the mean line-of-sight velocity dispersion
within the core, $\sigma_{c}$.
From Fig.~12 in Westfall et al.~(2006), one can see that it is
a generously taken value. 

\subsection{Escape velocity in the quasi-Newtonian MOND limit: Sculptor
as a reference case}

The effective gravitational mass of a isolated galaxy under MOND increases
linearly with $R$, and hence, any test particle is bound to the
potential. However, due to the nonlinearity of the MOND field equation, 
for a satellite system situated in an external gravitational
field, there exists a radius at which particles are stripped from the
dSph by the tidal field of the host galaxy (see \S \ref{sec:tides}).
For certain binary kick velocities, the LMXB system
could become unbound from the gravitational potential of the satellite.
Our aim in this Section is to estimate the kick velocity of a 
presupernova binary necessary to acquire the escape velocity, denoted by
$\vec{v}_{k}$. In Sculptor, the calculation greatly simplifies
because of the following reasons.
It can be seen that the internal acceleration $\sim 1.6\sigma^{2}/r_{c}
\sim 2\times 10^{-9}$ cm s$^{-2}$
felt by a star in the core of Sculptor 
is significantly smaller than $a_{0}$ and thus, for the dynamics of an
object in Sculptor the deep MOND regime would apply.
In addition, the parameter $\eta$
is $\lesssim 0.8$ for $V\gtrsim 185$ km s$^{-1}$. 
Since the internal acceleration goes as $\propto r$
at small radii and as $\lesssim r^{-1}$ at large radii, this value of $\eta$
points to domination of the external field throughout the dwarf galaxy
for $V\gtrsim 185$ km s$^{-1}$ (see also Milgrom 1995). 
At $V= 170$ km s$^{-1}$, we get $\eta\approx 1$, implying
it is in the borderline of isolation
with respect to the external acceleration field. 
Neither limit is valid in this case but the two limits should give
similar values for the escape speed.

A rough derivation of the escape speed follows; a more detailed
calculation is presented in Appendix \ref{sec:E1}.
Suppose that a certain dSph can be described
by a one-component King model (i.e.~mass follows light).
Then, the mass profile can be
parameterized by $W_{0}$, $\rho_{0}$ and $r_{c}$, with $\rho_{0}$
the total mass density at the center. In Newtonian
dynamics, $W_{0}$ is the depth of the potential in units of the
square velocity-dispersion parameter $\hat{\sigma}^{2}\equiv 
4\pi G \rho_{0} r_{c}^{2}/9$.
The escape velocity at any position is, therefore, smaller or equal
to the central escape velocity $\sqrt{2W_{0}\hat{\sigma}^{2}}$. 
For the Galactic dSphs, $W_{0}$ lies in the range $2<W_{0}<5$.

Consider now what happens under MOND theory.
If this galaxy is in the quasi-Newtonian regime and MOND 
has the ability of reproducing the dynamics of this galaxy
without dark matter, the depth of the potential under MOND
must be similar to that in the case of the Newtonian dark matter scenario.
Hence the central escape velocity is also $\sqrt{2W_{0}\hat{\sigma}^{2}}$
under MOND.  In the case of Sculptor
$W_{0}\approx 4$ (IH95; Walcher et al.~2003; Westfall et al.~2006). 
For this value of $W_{0}$, the velocity-dispersion parameter is
related to the observed velocity dispersion within the core through 
$\hat{\sigma}=1.16\sigma_{c}$ (see Fig.~4-11 in Binney \& Tremaine 1987).
Adopting $\sigma_{c}=8.8$ km s$^{-1}$ (see \S \ref{sec:sculptormain}),
the central escape velocity in Sculptor under MOND is
$\simeq \sqrt{10.8\sigma_{c}^2}=29$ km s$^{-1}$. 

The escape speed depends, of course, on radius. For instance,
the escape velocity at the core radius of
the same King model has been calculated to be $23$ km s$^{-1}$. 
Since the rms velocities of the progenitors in
the central parts of Sculptor are $\sim 8.8$ km s$^{-1}$, 
kick velocities of $\sqrt{23^{2}-8.8^2}\sim 21.5$ km s$^{-1}$ are enough 
for a LMXB formed at a core radius to be expelled. Consequently,
if the centers of
mass of all LMXBs have a recoil velocity of $21.5$ km s$^{-1}$, 
a fraction of them are retained and the rest becomes unbound,
depending on the location where the kick occurs. 
Since the fraction of mass within the core radius is
$35\%$, a significant fraction ($\sim 65\%$) of the new formed LMXBs 
will escape for a kick velocity of $21.5$ km s$^{-1}$ 
(a statistical approach to infer the mean kick velocity is given 
in Appendix \ref{sec:E1}, reaching the same result).

The systems that survive as binaries and become LMXB progenitors 
attain a system velocity of $180\pm 80$ km s$^{-1}$, if the distribution
of angles between the kick velocity and the orbital plane of the
presupernova binary is isotropic. If the range of kick directions
is restricted to a cone along the spin axis with an opening of $20^{\circ}$, 
LMXBs are launched with kick velocities of order
$20$--$100$ km s$^{-1}$ (e.g., Podsiadlowski et al.~2005). 
In this case and if the distribution of kick velocities were drawn randomly
in the range $20$--$100$ km s$^{-1}$, less than $2$ percent of the
LMXBs would be retained in Sculptor. The recent analysis of the Galactic 
pulsar proper-motion data by
Arzoumanian et al.~(2002) found evidence for a bimodal distribution,
with $\sim 40\%$ of the pulsars contained in a Maxwellian component 
with a dispersion of $\sim 90$ km s$^{-1}$,
and the remaining neutron stars in a Maxwellian component 
with a dispersion of $\sim 500$ km s$^{-1}$.
With this probability distribution, the retention fraction is
slightly larger but still less than $2.8\%$.

The presence of a heavy binary companion at the time of the
supernova explosion will make the retention more likely (e.g., Davies
\& Hansen 1998). 
Pfahl et al.~(2002) consider the inclusion of binaries
and conclude that the retention fraction is probably not
larger than several percent when they apply a single Maxwellian
fast kick mode at $200$ km s$^{-1}$ and a central escape speed of 
$50$ km s$^{-1}$.
Scaling down their results, this implies that the retention fraction
in Sculptor is likely not larger than a few percent even assuming
that the velocity kick distribution is described by
a slow Maxwellian mode at about $90$ km s$^{-1}$.
It is therefore puzzling to understand how Sculptor was able to
hold on to most of its X-ray binaries in MOND theory. 

One possibility is to invoke dark matter in MOND. Even if Sculptor 
obeys MOND, it can also contain a dark matter halo
more extended that the optical galaxy such that the kinematics
within the visible extent of Sculptor dwarf would not be greatly affected,
but will be very effective at holding on to the LMXBs in the dSph potential.
In this model, the MOND paradigm lowers the discrepancy between 
the binding mass and the baryonic mass but 
it still requires a total mass-to-light ratio 
$(M/L)_{M}\gtrsim 120$, where we have used Eq.~(\ref{eq:qnrel}) with
$(M/L)_{N}\gtrsim 600$ \citep{deh06}.
Again, the inclusion of a dark halo in MOND weakens its appeal.

\subsection{Compact early conditions}
\label{sec:compact}
It is simple to see that
if the stellar distribution was significantly more compact in
the past than it is at present, the probability of old LMXBs to be
bound to the potential increases. This is true for both Newtonian
and MOND dynamics, but applies only if the compact
dSph was subsequently heated to its current condition by a slow
adiabatic process. A dSph galaxy could have been more massive in the past
since mass-loss events 
can be caused by ram pressure stripping, supernovae explosions
and tidal stripping. However, since these mechanisms are thought to be
sudden and violent, it is very unlikely that they are able to produce
the desired effect. While \citet{rea05_01} have simulated the
dynamical effects of mass loss on the remaining stars and dark matter
in the Newtonian case, it is illustrative to see that, in fact, 
the velocity dispersion of baryons is not altered enough to consider 
mass loss as a promising alternative possibility to explain the
retention of observed LMXBs. In the recent simulations by Mayer et al.~(2007),
the formation of the darkest dSphs like Draco and Ursa Minor,
is the result of the transformation from a gas-rich dwarf to a dSph 
by repeated tidal shocks at pericenters when the gas is readily removed
by ram pressure stripping. Again, these impulsive shocks are not able
to change the stellar properties in an adiabatic way. In this scenario,
dSphs that fall into the Milky Way halo late, suffer little central
stripping and may have periodic bursts of star formation because
of funneling of this gas by a tidally-induced bar.

In the dark matter picture, there might exist another somewhat
speculative via to produce dynamical heating of the stellar
population. If the dark matter component is made up by massive black
holes, gravitational encounters with stars will produce a transfer
of energy between black holes and stars, in the attempt to reach
equipartition. \citet{jin05} have demonstrated that
the dynamical heating of an initially compact stellar distribution
might produce a remnant distribution similar to those in dSphs
for black hole masses in excess of $10^{5}$ M$_{\odot}$. 
LMXBs formed at early stages ($\gtrsim 3$ Gyr ago) 
could become bound to the potential well of the dSph galaxy.
However, this scenario is highly unlikely 
(Hernandez et al.~2004; S\'anchez-Salcedo \& Lora 2007).

\subsection{Comparison with the escape velocities
within the dark matter paradigm}
For a central line-of-sight velocity dispersion
of $6.2\pm 1.1$ km s$^{-1}$ as measured in Sculptor, 
the central mass-to-light ratio in the classical Newtonian
view was estimated to be $6$--$13$
(Queloz et al.~1995; Westfall et al.~2006). 
Given its large uncertainties,
this range of values is consistent or marginally larger than the
stellar mass-to-light ratio expected for an old stellar population
(Queloz et al.~1995). Nevertheless, its kinematics could be also
compatible with Sculptor being hosted by an extended subhalo.
In fact, a mass estimator for dark matter halos via the escape-velocity
argument was used by \citet{deh06}.
They construct diagnostic dark halo models with cumulative mass profiles
described by a simple analytic formula, designed to explore the generic
parameters required for retaining LMXBs in Sculptor. These halos have
different core radii, but all satisfying that the dark mass within the
visible galaxy, with a radius of $\sim 1.85$ kpc, be
$5\times 10^{7}$ M$_{\odot}$. 
They find that no reasonable dark matter
halo in Sculptor can retain all LMXBs with  
ejection velocities in the range 20-100 km s$^{-1}$. To hold on to LMXBs with
ejection velocities
of up to 60 km s$^{-1}$, a dark matter halo having a total radius of
upwards of 15 kpc is required.
LMXBs with lower ejection speeds are more easily retained and require
dark matter halos with more conservative parameters.

\begin{figure}
\plotone{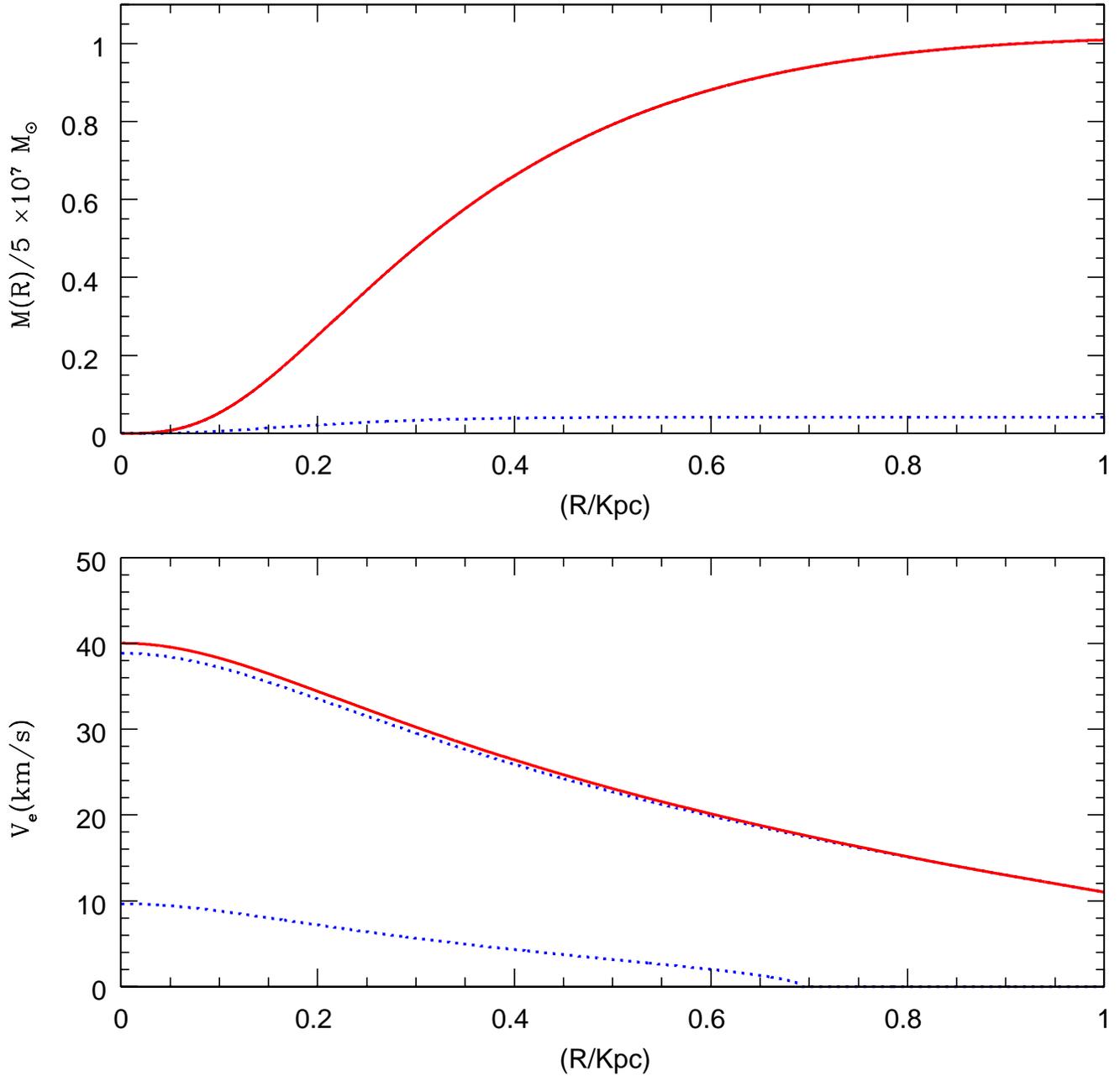}
  \caption{The upper panel shows a mass model for Sculptor constrained
to agree with observed total stellar content and scale length
(dotted curve), and a dark matter mass within $1.85$ kpc of
$5 \times 10 ^{7} M_{\odot}$, compatible with 
the stellar velocity dispersion within this region. 
The core radius of the dark halo is $\sim 0.3$ kpc. The lower panel
gives the contributions to the escape velocity at different radii,
from stars (lower dotted curve), dark halo (upper dotted curve),
and total (solid curve). The dynamics is entirely dominated by the
dark matter halo, which can only retain LMXBs having a low initial kick.}
  \label{fig:escape}
\end{figure}

In order to make the comparison with MOND as fair as possible,
we have explored the parameter space that ensures that
LMXBs are retained in Sculptor but using full
King models for the dark halo of Sculptor.  Consistent with the
findings of \citet{deh06}, we find that no realistic halo
model is sufficient to retain LMXBs having large ejection
velocities. Given that these ejection velocities are expected
in the range of 20-100 km s$^{-1}$ (e.g., Podsiadlowski et al.~2005),
and that the velocity dispersion in Sculptor is of only around 9 km s$^{-1}$,
this conclusion is not surprising. Actually, for the dark mass
limits within the visible galaxy coming from dynamical
studies, and imposing halo total radii smaller than 15 kpc,
little room for varying the King halo
parameters remains. For reasonable parameters of the dark halo,
escape velocities are always of order 30-40 km s$^{-1}$.
Figure \ref{fig:escape} 
shows one such mass model for Sculptor, with the upper panel
giving the cumulative mass profile
for the stars (also taken as a King distribution with a stellar
mass-to-light of $1.5$ and a concentration of $0.63$), and the total mass.
The dominance of the dark matter halo is dramatic,
especially at large radii,
although not extraordinary in comparison to inferences of large galaxies.
The bottom panel gives the
escape speed as a function of radius together with the components
due to the stars and dark matter.
We see that even at the very center,
the escape velocity is of only 40 km s$^{-1}$, and drops to 30 km s$^{-1}$ at a
distance of around 0.3 kpc. One should perhaps
expect the progenitors of LMXBs to have been located within the
central $0.165$ kpc of Sculptor, which defines
the core radius of the stellar populations. This leads to the
conclusion that standard dynamics
and a reasonable dark matter halo proposal for Sculptor can
explain the retention of LMXBs, only for those having started with a
relatively low $\leq 40$ km s$^{-1}$ initial kick. 
This is a more satisfactory situation than in the MOND case.
In order to keep
neutron stars with kick velocities $\gtrsim 50$ km s$^{-1}$ bound,
an extraordinarily extended and massive halo with a very large
core radius should be invoked.
Additionally, if a large fraction of neutron stars in dwarfs were formed 
in binary systems, 
the retention of LMXBs in Sculptor could be more plausible, but not trivial,
certainly not for all initial velocities
in the theoretical ejection range of velocities, at least not with
the restriction that the total $M/L\lesssim 35$ within $1.85$ kpc.
Therefore, the problem of
explaining the large retention of LMXBs is not exclusive of GCs
(e.g., Pfahl et al.~2002).

\section{Conclusions}
We have pushed the measurements of tidal radii in Galactic GCs
and dSph galaxies attempting to distinguish between dark matter and MOND. 
Except for Pal 14, the predicted tidal radii of GCs are systematically
larger than the observed nominal values. However, this is not
enough to rule out MOND because 
after including properly all the uncertainties,
we find that they are consistent with the observed values at a $1\sigma$ level.
The importance of the Galactic tide on the survival of dSph galaxies
has been also investigated. Assuming mass-to-light ratios compatible with
a naked stellar population, the present dSph galaxies preserve their
integrity except perhaps Sextans, which might be undergoing tidal
disruption.  We suggest that deep observations of
Sextans might provide a further test of MOND.

Based on the most recent data and with the updated values of $a_{0}$ 
and $M_{G}$,
the dynamical mass-to-light ratios inferred under Newtonian and MOND gravities
have been derived and compiled to see if new data
can definitively tell us what the law of gravity 
at small galactic scales is ($\sim 300$ pc).
Since the dynamical mass in MOND should be dominated by stars,
it is worrying that it requires mass-to-light ratios $\gtrsim 12$
for those dSph galaxies with
radially extended data right to the optical edge (Ursa Minor and Draco).
We warn that Willman 1, Coma Berenice, Ursa Major and Bo\"{o}tes may be also 
problematic for MOND.
In particular, preliminary estimates for Willman 1, Coma Berenice
and Ursa Major suggest  
$M/L\gtrsim 20$, at the $95 \%$ confidence level, 
even when possible outliers that
inflate the inferred velocity dispersion, and hence the mass-to-light
ratio, are omitted. That MOND requires a fraction of dark matter in 
clusters of galaxies is a well-established issue. 
Taken at face value, our results indicate that MOND needs also a dark
component in dSph galaxies with a mass fraction
much beyond that required in galaxy clusters, 
weakening the appeal of 
the MOND paradigm, although this fact by itself does not rule out
MOND as the gravity law at low accelerations.

An estimate of the mass of Sculptor has been given via
the escape-velocity argument; LMXBs are ideal probes of the total
mass as they should penetrate well outside the visible galaxy
due to their high recoil velocities $\gtrsim 20$ km s$^{-1}$.
In the standard dark matter scenario, LMXBs with velocities 
$\gtrsim 40$ km s$^{-1}$
are difficult to be retained for reasonable parameters of the dark halo.
Hence, the well-known retention problem of LMXBs in GCs
persists in dSphs. In the MONDian case,
we have calculated that the mean square kick velocity of a presupernova
binary to acquire the escape velocity is $\sim 24$ km s$^{-1}$ in Sculptor.
About $95\%$ of the formed LMXBs would have 
escaped and been stripped from the Sculptor galaxy.
In order for Sculptor to hold on to LMXBs with kick velocities
of $50$ km s$^{-1}$, a mass-to-light ratio $\gtrsim 120$ is required
even under MOND hypothesis, making it structurally different.
Therefore, either this is new evidence for the
necessity of an extended dark halo in a galaxy in MOND,
or the conventional thinking regarding neutron stars kicks must be modified.
Measuring the radial velocities of the observed LMXBs would provide
an obvious test of alternative gravities. High radial velocities 
$\gtrsim 30$ km s$^{-1}$ may confirm the presence of a very massive
dark matter halo much more extended than the stellar population.

Finally, one could argue that the large values of the $M/L$ in
MOND must be attributed to a problem with the velocity dispersion
measurements and their estimated uncertainties.
The required change in the velocity dispersions appears 
unlikely on purely statistical grounds and, in addition,
they do not act in a systematic manner in Draco and Ursa Minor
when compared to Sculptor.

In a MOND universe, data for several of these systems imply the presence 
of an extended dark matter halo. 
Although not a disproof of MOND, it does reduce the appeal of MOND 
significantly.

\acknowledgments
We would like to thank the anonymous referee for a careful reading of the
manuscript and useful suggestions which improved the presentation of our
final version. The authors acknowledge support from PAPIIT project
IN-114107-3.

\appendix
\section{MOND in the Quasi-Newtonian Limit: 
Virial Theorem} 
\label{sec:A1}

For a satellite galaxy embedded in a constant external field $g_{\rm ext}$,
the gravitational field equation
that governs the kinematics of the stellar component in the
quasi-Newtonian limit is 
\begin{equation}
\left(\nabla^{2}+\frac{\partial^{2}}{\partial z^2}\right)\Phi=
\frac{a_{0}}{g_{\rm ext}}4\pi G \rho.
\label{eq:PoissonMOND}
\end{equation}
The solution of this modified Poisson equation 
for a distribution of mass $\rho(\vec{r})$ is given by
\begin{equation}
\Phi(\vec{r})=-\left(\frac{Ga_{0}}{g_{\rm ext}}\right)
\int\frac{\rho(\vec{r}')}{\sqrt{2(x'-x)^{2}+2(y'-y)^{2}+(z'-z)^{2}}}d^{3}
\vec{r}'.
\label{eq:Phigeneral}
\end{equation}

As usual, the virial theorem for a stationary system can be derived after
integration over an arbitrary volume $\tilde{V}$:
\begin{equation}
\int_{\tilde{V}} \rho v^{2} d^{3}\vec{r}=
\int_{\tilde{V}}\rho \vec{r}\cdot\vec{\nabla}\Phi
\,\,d^{3}\vec{r}.
\label{eq:virial1}
\end{equation}
Substituting Eq.~(\ref{eq:Phigeneral}) into Eq.~(\ref{eq:virial1}), one
gets:
\begin{equation}
2K+W=0,
\label{eq:virialgen}
\end{equation} 
where $K$ is the total kinetic energy of the system and $W$ is the system's
total potential energy:
\begin{equation}
W\equiv\frac{1}{2}\int\rho(\vec{r}) \Phi(\vec{r})\,\, d^{3}\vec{r}.
\label{eq:potentialenergy}
\end{equation} 
We have seen that the virial theorem has the same form than in the Newtonian
case, with $\Phi$ as given by Eq.~(\ref{eq:Phigeneral}).

\section{Potential energy of a homogeneous sphere in the
quasi-linear limit}
\label{sec:B1}
We wish to calculate the potential energy, $W$, of a homogeneous sphere
of constant density $\rho$ and radius $r_{0}$ in the quasi-linear MONDian
limit. According to Eq.~(\ref{eq:potentialenergy}),
we need $\Phi$ in order to evaluate $W$. The field equation 
(\ref{eq:PoissonMOND}),
can be transformed into the standard Poisson equation by making
the substitution $z'=z/\sqrt{2}$:
\begin{equation}
\tilde{\nabla}^{2}\Phi'=\frac{a_{0}}{g_{\rm ext}}4\pi G\rho(x,y,\sqrt{2}z'),
\end{equation}
where $\Phi'=\Phi'(x,y,z')$ and $\tilde{\nabla}^{2}=\partial^{2}/\partial x^{2}+
\partial^{2}/\partial y^{2}+\partial^{2}/\partial z'^{2}$. The
relationship between the potentials is $\Phi(x,y,z)=\Phi'(x,y,z/\sqrt{2})$.
For a sphere of constant density, $\Phi'$ satisfies:
\begin{equation}
\tilde{\nabla}^{2}\Phi'=\frac{a_{0}}{g_{\rm ext}} 4\pi G \rho',
\end{equation}
where $\rho'=\rho$ within the ellipsoidal body bounded by the
surface $r_{0}^{2}=R^{2}+2z'^{2}$, which is an oblate ellipsoid 
with eccentricity $e=0.71$. 
We are now in a position to calculate $W'=1/2\int \rho'\Phi' d^{3}\vec{r}'$ 
taking advantage of the theory of homoeoids 
(e.g., Roberts 1962; Binney \& Tremaine 1987):
\begin{equation}
W'=-\frac{8}{15\sqrt{2}}\pi^{2}\left(\frac{a_{0}}{g_{\rm ext}}\right)G
\rho^{2}r_{0}^{5}I(e),
\label{eq:energyprime}
\end{equation}
with $I(e)$ an analytic function of the eccentricity as given in Table 2-2 
of \citet{bin87}, being $I=1.57$ for $e=0.71$. 

Finally, we can obtain $W$, $W=1/2\int \rho\Phi d^{3}\vec{r}$, in
terms of $W'$ by expressing $W$ in terms of the new variable of integration 
$z\rightarrow z'$.  This leads to $W=\sqrt{2}W'$. Substituting the
value of $W'$ from Eq.~(\ref{eq:energyprime}) as a function of the
total mass $M=4\pi \rho r_{0}^{3}/3$, we find 
\begin{equation}
W=-0.47\left(\frac{a_{0}}{g_{\rm ext}}\right)\frac{G M^{2}}{r_{0}}.
\end{equation}
If the MOND dilation along the $(x,y)$-directions were ignored and
the potential were assumed to be isotropic rather than anisotropic,
the potential energy would be $-0.6(a_{0}/g_{\rm ext})GM^{2}/r_{0}$. 
Hence, the error made using the isotropic approximation is
$28\%$, for a homogeneous sphere.

\section{Dynamical mass-to-light ratios of dSphs}
\label{sec:D1}
The closest and faintest dwarf in our sample is Willman 1
at a distance of 40 kpc and with a luminosity of $\sim 855$ L$_{\odot}$ 
(Martin et al.~2007).
For the reported core radius of $16.6$ pc, this galaxy may be
treated as isolated for internal velocity dispersions $\gtrsim 2.7$ km s$^{-1}$,
corresponding to masses of $8.5\times 10^{4}$ M$_{\odot}$. Therefore,
for realistic mass-to-light ratios of a stellar population, Willman 1
is in the quasi-linear regime.
The expected velocity dispersion is:
\begin{equation}
\sigma=0.36 \sqrt{\frac{a_{0}}{g_{\rm ext}}}\sqrt{\frac{GM}{r_{hp}}},
\label{eq:bau}
\end{equation}
where $r_{hp}$ is the projected half-mass radius (e.g., Baumgardt et al.~2005).
Suppose for a moment that the mass-to-light ratio is $\sim 3$.
Under this assumption, Eq.~(\ref{eq:bau}) implies 
$\sigma=0.53\pm 0.1$ km s$^{-1}$, where $r_{hb}\approx 21\pm 7$ pc was used.
By contrast, Martin et al.~(2007) measure a radial velocity
dispersion of $4.3\pm^{2.3}_{1.3}$ km s$^{-1}$ and higher than $2.1$
km s$^{-1}$ at the $99\%$ confidence limit. Therefore, if Willman 1
does not contain dark matter, it is already gravitationally unbound
even under MOND, and will disperse in a few crossing-times, i.e. in a few
$\sim 10^{7}$ yr. This is at odds with the photometric data that do
not show tidal tails or visible increase of the velocity dispersion
with distance (Martin et al.~2007). Thus it would seem that
Willman 1 needs a dark matter halo even in MOND.

In order for Willman 1 to be in virial equilibrium
with a velocity dispersion of $\sim 4.3$ km s$^{-1}$, we require
a mass-to-light ratio between $370$ and $550$. This estimate was derived
from the Newtonian values reported in Martin et al.~(2007) but
noting that since the system is not in deep MOND in this case
(the internal accelerations are of the order of $a_{0}$), 
the MOND $M/L$ value is only a factor of $\sim 1.3$ smaller than the
Newtonian values \citep{mil94}. Adopting the lower limit of 
$2.1$ km s$^{-1}$
at the $99\%$ confidence level, Willman 1 lies in the quasi-linear
deep-MOND regime and then $(M/L)_{M}>30$.

For the new Galactic dSph candidate found in the
constellation of Bo\"{o}tes, Martin et al.~(2007) have 
derived a velocity dispersion $6.5\pm ^{2.1}_{1.3}$ km s$^{-1}$ from 24
members of the Bo\"{o}tes dwarf. For the adopted distance of $60$ kpc,
this galaxy has a core radius of $\sim 225$ pc and is well in 
the quasi-Newtonian regime. Adopting a total luminosity of
$1.6\times 10^{4}$ L$_{\odot}$ to $8.6\times 10^{4}$ L$_{\odot}$,
implies $(M/L)_{0,M}\sim 120$ to $23$.
Given the uncertainties in the velocity dispersion, 
$(M/L)_{0,M}\gtrsim 15$ at the $68\%$ confidence.
We must stress here that the magnitude and associated errors 
in the velocity dispersion 
and structural parameters of Bo\"{o}tes and Willman 1
are currently crude estimates.

For Ursa Minor, Wilkinson et al.~(2004) derived a 
line-of-sight rms dispersion of 
$12$ km s$^{-1}$, with $\sigma>9.5$ km s$^{-1}$ at the $95\%$ confidence
level. A robust estimate of the total mass can be found from
Eq.~(\ref{eq:isorel}) provided that the velocity distribution is isotropic.
For a value of $\sigma=12$ km s$^{-1}$ we get $(M/L)_{M}=60$ even taking
the luminosity derived by Palma et al.~(2003), while 
$(M/L)_{M}>23$ at the $95\%$ confidence level.

Based on the Newtonian value $(M/L)_{N}=9$ and in the quasi-Newtonian limit 
(Eq.~\ref{eq:qnrel}) with $g_{\rm ext}=V^{2}/D$ and $V=170$ km s$^{-1}$,
the global $(M/L)_{M}$ in Sculptor is $1.0\pm 0.5$,  
for the observed central line-of-sight velocity dispersion
$6.2\pm 1.1$ km s$^{-1}$ (Queloz et al.~1995; Westfall et al.~2006).
The real global $M/L$ ratios are probably a factor $\times 2$ larger 
because the velocity dispersion shows a rise, leading to a mean
velocity dispersion in the core of $8.8\pm 0.6$ km s$^{-1}$ 
(Westfall et al.~2006). 
In any case, the mass-to-light ratios in MOND are perfectly consistent
with an old stellar population.
In a Newtonian dark matter view, a flat velocity-dispersion profile
well beyond the several core radius, as occurs in Sculptor,
is interpreted as the presence of a halo
that begins to dominate dynamics increasingly outwards.
However, in the MOND quasi-Newtonian limit, such a behavior for 
the velocity dispersion, even if the
actual values of the stellar mass-to-light ratios remain within
plausible ranges, would imply a radial
change in the intrinsic properties of the stars,
something hard to explain in such small systems, or that the dwarf
galaxy is being heated by Galactic tides.

Draco and Fornax were studied by \citet{lok01, lok02}.
The $(M/L)_{M}$ values for isotropic models after updating by
$a_{0}$ and the luminosities are $29\pm 9$ and $2.2\pm 0.4$, respectively. 
The mass of Leo II and Leo I have been estimated with 
Eq.~(\ref{eq:isorel})
using the mean velocity dispersion $6.6\pm 0.7$ km s$^{-1}$ 
(Koch et al.~2007b) and $9.9 \pm 1.5$ km s$^{-1}$ (Koch et al.~2007a),
respectively.
For Carina, we estimate $(M/L)_{M}$ with Eq.~(\ref{eq:centralMOND}) 
adopting the central 
value $6.97\pm 0.65$ km s$^{-1}$ (Mu\~{n}oz et al.~2006a).

For Sextans, the two-component models by 
Kleyna et al.~(2004) suggest $(M/L)_{N}$ between $60$--$300$ within
$1$ kpc, being the exact value sensitive to the form assumed for the outer
fall-off in the velocity dispersion profile. 
Based on these Newtonian values, assuming that the external field
is dominant in Sextans, a $(M/L)_{M}$ 
of $\sim 7.5$--$36$ was found when adopting a distance of $86$ kpc.
Although $M/L$ of GCs have a $3\sigma$ range of $1$--$4.3$
(Feltzing et al.~1999; Parmentier \& Gilmore 2001), 
values as high as $7$--$8$ cannot be ruled out for a stellar old population
(Queloz et al.~1995; Romanowsky et al.~2003).
Hence, a mass-to-light ratio of $\sim 8$ seems to be compatible 
with its kinematics and might be associated to a naked stellar old component.

In the case of the recently discovered Ursa Major
dSph, for a central velocity dispersion of $7.6$ km s$^{-1}$ 
and a core radius $r_{c}\sim 200$ pc (Simon \& Geha 2007),
we obtain a central mass-to-light ratio of $135$ in the isolated limit.
For the value reported by Martin et al.~(2007), 
$\sigma=11.9\pm^{3.5}_{2.3}$ km s$^{-1}$, the mass-to-light ratio is even
larger.
The velocity dispersion is in fact greater than $6.0$ km s$^{-1}$ at a $95\%$
confidence level. Adopting this lower value, 
Ursa Major lies in the quasi-Newtonian
regime and, then, $(M/L)_{0,M}>52$ with $95\%$ confidence. 
The magnitude of Ursa Major is currently a crude estimate. 
Improving the data on this galaxy is key to exploring how MOND
behaves at the smallest galactic scales ($\sim 300$ pc).

Measurements of the velocity dispersions for Coma Berenices, Hercules
and Canes Venatici II were reported in \citet{sim07}.
They measure the velocities of $59$ stars in Coma Berenice,
obtaining a velocity dispersion $4.6\pm 0.8$ km s$^{-1}$.
Equation (\ref{eq:centralMOND}) with $r_{c}=41$ pc and $(M/L)_{N}=448$ 
implies a MOND mass-to-light ratio of $105$. 
Hercules is likely dominated by the external acceleration because
$\eta\simeq 0.91$ for $r_{c}=205$ pc and assuming a distance of $140$ kpc. 
For a velocity dispersion $5.1\pm 0.9$ km s$^{-1}$ and $(M/L)_{N}=332$,
we obtain $(M/L)_{0,M}=26\pm^{11}_{9}$. Finally, the ultra-faint Canes
Venatici II is located at $150$ kpc. We have used $r_{c}=84$ pc
and $\sigma=4.6\pm 1$ km s$^{-1}$ to infer 
$(M/L)_{0,M}=38.5\pm^{46}_{24}$.

Since the global luminosity is less certain than the central
luminosity, we can use the central mass-to-light estimators  
$(M/L)_{0,M}$ as a check of the robustness of the estimates, 
for Draco and Ursa Minor.  Assuming that mass follows
light, Odenkirchen et al.~(2001) obtain $(M/L)_{N}=146$ for Draco. If we take
this value as representative of the Newtonian value within the optical
radius of Draco, and for a central velocity dispersion $\sigma_{0}=8$ 
km s$^{-1}$ (Lokas et al.~2006) and core radius of $180$ pc, 
Eq.~(\ref{eq:centralMOND}) gives $(M/L)_{0,M}=24$ 
\footnote{\citet{mil95} obtained a central $M/L=4$, using
$\sigma_{0}=9.2$ km s$^{-1}$ and $a_{0}=2\times 10^{-8}$ cm s$^{-2}$.}.
In the case of Ursa Minor, the central Newtonian
value is $100$ after updating the quoted value of \citet{mat98}
with the latest determination of $\sigma_{0}=12$ km s$^{-1}$
(Wilkinson et al.~2004), instead
of $9.3$ km s$^{-1}$ adopted in \citet{mat98}. From Eq.~(\ref{eq:centralMOND})
with $r_{c}=200$ pc, we get $(M/L)_{0,M}=33$.
These central values of $(M/L)_{0,M}$ are therefore in agreement with  
the global estimates.

\section{Mean square kick velocity to escape}
\label{sec:E1}
Consider a progenitor system of a LMXB, which is embedded in a galaxy 
that lies in the quasi-Newtonian regime in MOND.
If the progenitor system is moving at velocity $\vec{v}$ and the supernova
explosion occurs at $\vec{r}$, then the minimum kick velocity
for the new formed LMXB to become gravitationally unbound, 
$\vec{v}_{k}$, satisfies:
\begin{equation}
\frac{1}{2}\left(\vec{v}+\vec{v}_{k}\right)^{2}+\Phi(\vec{r})
-\frac{1}{2} |\vec{\Omega}\times \vec{r}| ^{2}=
\Phi(\vec{r}_{t})-\frac{1}{2} |\vec{\Omega}\times \vec{r}_{t}| ^{2},
\end{equation} 
where $\vec{\Omega}$ is the angular velocity of the dwarf satellite around
its parent galaxy, $\vec{r}_{t}=(0,0,r_{t})$ and $r_{t}$ the tidal radius along
the line pointing the centers of the galaxies. 
The mean square kick velocity for stars to escape is:
\begin{equation}
\left< v_{k}^{2}\right>=-\left<v^{2}\right>-
2\left(\left<\Phi(\vec{r})\right>-\Phi(\vec{r}_{t})\right)
+<|\vec{\Omega}\times \vec{r}| ^{2}>
-|\vec{\Omega}\times \vec{r}_{t}| ^{2},
\label{eq:avcoiling}
\end{equation} 
where the brackets represent the mass average: 
\begin{equation}
\left<a\right>\equiv \frac{\int \rho(\vec{r}) a d^{3}\vec{r}}
{\int \rho(\vec{r}) d^{3}\vec{r}}
=\frac{1}{M_{dw}}\int \rho(\vec{r}) a d^{3}\vec{r}.
\end{equation}
In Eq.~(\ref{eq:avcoiling}) we used $\left<\vec{v}\cdot \vec{v}_{k}\right>=0$ 
for an isotropic distribution of kick directions.

In the quasi-Newtonian MOND regime,
Equation (\ref{eq:avcoiling}) can be expressed in terms of 
observable quantities as follows. First, the quantity 
$\left< |\vec{\Omega}\times \vec{r}| ^{2} \right>$
is exactly $2\Omega^{2}\left< r^{2}\right>/3$ for a 
spherical distribution of matter. 
Corrections for non-sphericity of the distribution in the central parts
of the dwarf satellite were found to be unimportant as far as 
$\left<v_{k}^{2}\right>$ concerns.
Ignoring the variation of the external gravitational field in the
vicinity of the satellite, $\left< v_{k}^{2}\right>$ can be written:
\begin{equation}
\left< v_{k}^{2}\right>=-\frac{2K}{M_{dw}}-
\frac{4W}{M_{dw}}+2\Phi(\vec{r}_{t})
+\Omega^{2}\left(\frac{2}{3}\left<r^{2}\right>-r_{t}^{2}\right),
\end{equation} 
where $K$ is the total kinetic energy of the stellar system and $W$ the
total potential energy\footnote{Note that the concept
of total potential energy is not well-defined in an isolated system
but it is meaningful in the quasi-Newtonian field limit.} 
(Appendix \ref{sec:A1}). 
It is easy to show that, independently of the concentration, 
$\left<r^{2}\right><r_{t}^2/5$ in any King model.
Using the virial theorem in the quasi-linear regime 
(see Eq.~\ref{eq:virialgen}),
with $K=3M_{dw}\sigma^{2}/2$, $\sigma$ the global one-dimensional
velocity dispersion,
$\Omega=V/D$ and
approximating $\Phi(\vec{r}_{t})\approx -G(a_{0}/g_{\rm ext})M_{dw}/r_{t}$,
it holds that:
\begin{equation}
\left< v_{k}^{2}\right>\leq 9\sigma^{2}-2\frac{a_{0}}{g_{\rm ext}}\frac{GM_{dw}}{r_{t}}
-\frac{13}{15}\left(\frac{r_{t}}{D}\right)^{2}V^{2}.
\end{equation}
The second term of the RHS can be expressed in terms of $\sigma$
exploiting again the virial theorem and noting that the potential
energy in a King model with $W_{0}=4$ is $|W|=(9/5)GM_{dw}^{2}/r_{t}$ in
Newtonian dynamics \citep{kin66}, which translates into an upper limit in MOND
due to the dilation factor (Eq.~\ref{eq:Phigeneral}), i.e.~
$|W|<(9/5)G(a_{0}/g_{\rm ext})M_{dw}^{2}/r_{t}$. Consequently,
\begin{equation}
\left< v_{k}^{2}\right>\leq 8\sigma^{2}
-\frac{13}{15}\left(\frac{r_{t}}{D}\right)^{2}V^{2}.
\end{equation}
 
For the parameters of Sculptor: $\sigma \simeq 0.95 \sigma_{c}=8.4$ 
km s$^{-1}$, $r_{t}=1.8$ kpc, $V=170$--$185$ km s$^{-1}$, $D\simeq 79$ kpc, 
we obtain $\left<v_{k}^{2}\right>^{1/2}\lesssim 23.5$ km s$^{-1}$.

\clearpage

\begin{table*}
\begin{minipage}{156mm}
\caption[]{A summary of relevant parameters for the closest dSph galaxies.
The second (and third) row for some of the objects gives the values
reported by other authors.  }
\vspace{0.01cm}
\small
\begin{tabular}{c c c c c c c c l}\hline
{Galaxy} & {L} & {$r_{t}$} & {$(M/L)_{N}$\tablenotemark{a}} & Comp.\tablenotemark{b} & {$F_{N}$} &
{$(M/L)_{M}$\tablenotemark{c}} & { $F_{M}$} & References for \\
{}&$10^{5}$L$_{\odot}$ &arcmin & & & & & &($L,r_{t},(M/L)_{N}$) \\

\hline
Willman 1 & 0.0085 & $\sim 7.4$ & 500 & 1 & 0.25 & $\gtrsim 30(95\%)$ 
& 0.36 & M07, M07, M07\\  
Coma Ber. & 0.027 & --- & $448\pm^{169}_{142}$ & 1 & --- & $\gtrsim 46(95\%)$
& --- & B06, ---, S07\\
Bo\"{o}tes & 0.86 & --- & 130 & 1 & --- & $23\pm^{17}_{8}$ & --- & M06b,---, M07\\
Ursa Minor   & 2.9  & $50.6\pm 3.6$  
& $79\pm 39$ & 1 & 0.55  & $12\pm 6$  & 0.51  & M98, IH95, M98   \\
  & 5.8   & $77.9\pm 8.9$  
&  $40\pm 20$ &  1 & 0.85  &  $6\pm 3$  &  0.77 & P03, P03, P03    \\
 & 5.8  & $\sim 34$  
& $340\pm 240 $ & 2 & 0.19  & $\gtrsim 23 (95\%)$  
& 0.16  & P03, W04, W04   \\
Sculptor   & 14   & $79.6\pm 3.3$  & $9\pm 6$ & 1 
& 1.13  & $1.0\pm 0.5$  & 1.10 &
IH95, We06, Q95  \\
   & 14   & $\sim 60$  & $9\pm 6$ & 1 & 0.86  & $1.0\pm 0.5$  & 0.83 &
IH95, We06, Q95  \\
Draco   & 2.6   & $28.3\pm 2.4$  & $\gtrsim 240$ & 2
&  0.24  & $\gtrsim 13 (95\%)$  & 0.28 & M98, IH95, K01  \\
& 2.4   & $\gtrsim 40.1$  & $140\pm 40$ &  1 
&0.42  & $\gtrsim 12 (95\%)$  & 0.40
& O01, O01, O01 \\
Sextans   & 5.0   & $160\pm 50$  & $107\pm 72$ &  1 &
1.40  & $22\pm 15$  & 1.05 
& M98, IH95, IH95   \\
Carina   & 4.3   & $28.8\pm 3.6$  & $43\pm^{53}_{19}$ &  1 &
0.42  & $4.5\pm 0.9$  & 0.35
& M98, IH95, M06a   \\
Ursa Major   & $0.15$   & ---  & $1024\pm 636 $ & 1 & ---  & $\gtrsim 52 (95\%)$  & --- 
& B06,---, S07   \\
Fornax   & 155   & $71.1\pm 4.0$  & $15\pm 10$ &  2 
& 0.46  & $2.2\pm 0.4$  & 0.33
& M98, IH95, Wa06  \\
  & 155   &  $90$  & $15\pm 10$ &  2& 0.58  & $2.2\pm 0.4$
& 0.42 & M98, W03, Wa06   \\
Hercules & 0.21 & --- & $332\pm^{127}_{106}$ & 1 & --- & $26\pm^{11}_{9}$ &
--- & B07, ---, S07 \\
Canes Vn II & 0.071 & --- & $336\pm^{162}_{130}$ & 1 & --- & $38.5\pm^{46}_{24}$
& --- & B07, ---, S07\\
Leo II  & 7.5   & $8.7\pm 0.9$  & $35.6\pm^{8.8}_{7.7} $ &  1 & 
0.19  & $4.3\pm 2.1$  & 0.09 & M98\tablenotemark{\ast}, IH95, K07b   \\
Leo I  & 47.9   & $12.6\pm 1.5$  & $17\pm 4$ &  1 & 
0.12  & $3.4\pm ^{2.5}_{1.5}$  & 0.07 & M98, IH95, K07a   \\
   & 76   & $13.4\pm 0.7$  & $5.3\pm 2.6$ &  1 & 
0.13  & $2.1\pm^{1.5}_{1.0}$  & 0.07 & S06, S06, S06   \\

\hline
\end{tabular}

\tablenotetext{a}{Mass-to-light ratios for the
Galactic dSphs with determinations of their internal kinematics.  
The quoted $1\sigma$ error bars do
not include the associated uncertainties in luminosity or distance. In galaxies
in which the reported value of the luminosity varies significantly 
between different authors, we have added a second row.
Obviously, in order to have low values of the $M/L$'s,
the larger values of the luminosity should be taken.}

\tablenotetext{b}{We indicate the mass model used. $``1"$ indicates one-single King 
component models --with or without dark matter. If the velocity dispersion
profile is available, the assumption that mass-follows-light can
be relaxed and more reliable models can be obtained. $``2"$ refers to
models in which dark matter and stars follow different
profiles ($2$-component models).}

\tablenotetext{c}{Confidence levels are given in parenthesis.}

\tablenotetext{\ast}{Updated values using $D=233$ kpc (Bellazzini et al.~2005).}

\tablecomments{Legenda of the acronyms: 
B06: Belokurov et al.~2006;
B07: Belokurov et al.~2007;
IH95: Irwin \& Hatzidimitriou 1995; 
K01: Kleyna et al.~2001;
K05: Kleyna et al.~2005; 
K07a: Koch et al.~2007a;
K07b: Koch et al.~2007b;
M98: Mateo 1998;
M06a: Mu\~{n}oz et al.~2006a; 
M06b: Mu\~{n}oz et al.~2006b; 
M07: Martin et al.~2007;
O01: Odenkirchen et al.~2001;
P03: Palma et al.~2003;
Q95: Queloz et al.~1995; 
S06: Sohn et al.~2006;
S07: Simon \& Geha 2007;
W03: Walcher et al.~2003; 
W04: Wilkinson et al.~2004; 
Wa06: Walker et al.~2006; 
We06: Westfall et al.~2006. 
}

\label{table:parameters}
\end{minipage}
\end{table*}

\clearpage

\clearpage

\clearpage

\end{document}